\documentstyle[twocolumn,prl,aps]{revtex}
\preprint {imsc/97/07/32}
\begin{document}
\draft
\title{ Geometry of N-particle clusters in two-dimensions: Some 
exact results}

\author{G. Date and M. V. N. Murthy}

\address
{The Institute of Mathematical Sciences, Madras 600 113, India.\\
}
\date{\today}
\maketitle
\begin{abstract}

We report on a study of a finite system of classical confined particles in
two-dimensions in the presence of a uniform magnetic field and interacting
via a two-body repulsive potential. We develop a simple analytical method
of analysis to obtain ground state energies and configurations.  We prove
analytically the minimum energy configurations are independent of the
nature of two body interaction and the magnetic field. In particular we
prove that the first transition from a single shell occurs when the number
of particles changes from five to six. These results are exact. 

\end{abstract}


\narrowtext

There has been considerable progress in recent years in the study of
electrons in quasi-two-dimensional systems both experimentally and
theoretically\cite{tapash}.  There are several examples of these systems
but the most important one from our point of view is that of electrons in
a quantum dot.  There have been several studies on the ordering and
transitions of charged particles in
two-dimensions\cite{calinon,lozovik,maksym,bedanov,rossler}. In a recent
Monte Carlo study, Bedanov and Peeters\cite{bedanov}(see also Bolten and
Rossler \cite{rossler}) have analysed the classical ground state of a
system of confined charged particles interacting through Coulomb
interaction. By minimising the classical energy, they obtain
numerically the shell structure in a cluster of N-particles. They have
systematically listed the shell structure in a ``Mendeleev" table for
$N\le 52$, and for a few large clusters. Similar results are available
also for logarithmic two-body interaction\cite{calinon}.  

In this paper we consider such a cluster in which the repulsive two-body
potential is either a power-law( or logarithmic). We devise a simple
analytical method to obtain the classical ground state energy in {\it two}
steps. First we minimise the energy for a {\it fixed} total angular
momentum (which is conserved), $J$ and then minimise this energy with
respect to $J$. This has two advantages. It reflects the quantum
degeneracy of the lowest Landau level for electrons in a uniform magnetic
field in the absence of any interaction, at the classical level itself and
secondly it allows one to do the second step minimisation over the
quantised values of the angular momentum. Here however we restrict
ourselves to classical analysis only and derive some exact results
analytically. 

In particular we show that: a) the configurations minimising the energy
are {\it independent} of the parameters of the Hamiltonian (eg magnetic
field) so long as the repulsive two body potential falls off as a
power-law (or vary logarithmically) with the relative distance. Only the
overall length scale is sensitive to these details. 2) Two special
configurations in which all the N particles are on a circle (referred to
as $\bigcirc$) and the one in which N - 1 are on a circle with one at the
center (referred to as $\bigodot$ ) are always (local) minimum energy
configurations.  We give exact analytical expressions for the
corresponding minimum energy for all $N$. The $\bigcirc$ has lower energy
for $N \le 5$ while $\bigodot$ has lower energy for $N \ge 6$. This
geometric transition is the first one to occur and is {\it independent} of
the precise form of the repulsive interaction. 3) While it is known {\it
numerically} that for $N \ge 9$ and for Coulomb
potential\cite{bedanov,rossler} the minimum energy configurations exhibit
approximate multi shell structure, the special configurations provide an
upper bound on the minimum energy for a whole class of interactions that
we consider here.  

The classical system we are interested consists of $N$ particles confined 
in an oscillator potential in a uniform magnetic field and interacting 
via a two body interaction. The Hamiltonian of such a system of 
particles given by,
\begin{equation}
H = \sum_{i=1}^{N}\left[\frac{(\vec p_i+\vec a_i)^2}{2m} 
+\frac{1}{2}m\omega^2 
r_i^2 \right] + \beta\sum_{i,j(\ne i)} \frac{1}{(r_{ij}^2)^{\nu}},
\end{equation}
where $\vec r_i$ and $\vec p_i$ denote the position and momentum vectors 
of the i-th particle. The vector potential, for a uniform magnetic field 
is given by,
\begin{equation}
(a_i)_x = -\omega_L y_i~~; ~~~~~~~~~(a_i)_y =~\omega_L x_i ,
\end{equation}
where $\omega_L$ is the Larmor frequency and $\vec r_{ij} =\vec r_i - 
\vec r_j$. The power $\nu$ (positive) is kept arbitrary. 
In what follows we also comment on case when the 
two-body interaction is of the form $ -\beta\sum_{i,j(\ne
i)}\log(r_{ij}^2/\rho^2)$ which is repulsive for $r_{ij}^2 <
\rho^2$. This may be closer to the real situation  as  
also the $\nu=1/2$ (Coulomb) in the case of electrons in a quantum 
dot\cite{zhang}. 
Recently, the $\nu=1$ case has also been analysed in detail\cite{johnson}. The 
Hamiltonian can be written in terms of dimensionless units by 
introducing a length scale $l= \sqrt{(\hbar/(m\omega)}$ which is the 
basic oscillator length. All distances are measured in terms of this 
basic length unit. Note that the $\hbar$ is introduced only as a 
convenience so that the energy is measured in units of $\hbar\omega$ and 
does not have any other significance as in the quantum case.  The 
momenta are measured in units 
of $\hbar/l$. 

The new Hamiltonian in these scaled units, but keeping the 
same notation,  may be written as 
\begin{equation}
\frac{H}{\hbar\omega} = \sum_{i=1}^{N}\left[\frac{\vec p_i^{~2}}{2} 
+\frac{1}{2}(1+\alpha^2) r_i^2 + \alpha j_i\right] + g\sum_{i,j(\ne i)} 
\frac{1}{(r_{ij}^2)^{\nu}}, 
\end{equation}
where $j_i = \vec r_i \times \vec p_i$, $\alpha = 
\frac{\omega_L}{\omega}$ and $g = \frac{\beta}{\hbar\omega} (l)^{2\nu}$.
Unless otherwise mentioned the summations run from 1 to $N$ hereafter.
While the original coupling constant $\beta$ was dimensional the new 
coupling constant $g$ is dimensionless. Hereafter we assume all the 
energies are measured in units of $\hbar\omega$ and do not write the 
units explicitly. 

For the first step of the minimisation we introduce the function,
\begin{equation} 
F = H+\lambda(J - \sum_i j_i),
\end{equation}
where $j_i$ are the single particle 
angular momenta and $\lambda$ is the Lagrange multiplier which enforces
the constraint $J = \sum_i j_i$. 
Setting 
$\delta F=0$, gives the necessary equations to determine the 
equilibrium configuration in the phase space,
\begin{eqnarray}
p_{ix} &=&~(\alpha+\lambda)y_i,   \label{eqpx}\\
p_{iy} &=&-(\alpha+\lambda)x_i,   \label{eqpy}\\  
(1-\lambda^2-2\alpha\lambda)\vec r_i &=&  4g\nu  \sum_{j(\ne i)} \frac{\vec 
r_{ij}}{(r_{ij}^2)^{\nu+1}}. \label{eqr}
\end{eqnarray}
These are the basic set of equations. Any solution to this set of
equations describes an equilibrium configuration but not necessarily
the one with the minimum energy for the given $J$. The case of logarithmic
interaction is obtained by simply setting the power $\nu=0$ and
by setting the prefactor to $4g$ instead of $4g\nu$ in
eq.(\ref{eqr}). With this proviso, unless otherwise mentioned, 
all subsequent equations also reproduce the log case. 

First we present a qualitative but a  general analysis of these basic
set of 
equations. To make the analysis simple, we introduce an auxiliary variable,
\begin{equation}
\phi = \sum_i r_i^2 . \label{phidef}
\end{equation}
The total angular momentum may now be written in terms of this
auxiliary variable as
\begin{equation}
J = \sum_i \vec r_i \times \vec p_i = -(\alpha+\lambda)\phi,\label{eqj} 
\end{equation}    
where we have made use of eqs.(\ref{eqpx},\ref{eqpy}).
It is convenient to express $ \vec r_i = R \vec s_i$ with $R$ being a
common scale factor which may be taken to be the radius of the farthest
particle, say the $N^{th}$ one.
Therefore 
\begin{equation}
\phi = R^2[\sum_{i=1}^{N-1} s_i^2 +1] \equiv R^2 \tilde \phi
\end{equation}
since $s_N^2 = 1$.
Using eq.(\ref{eqr}) and eliminating $\lambda$ dependence using
eq.(\ref{eqj}) we have, 
\begin{equation}
\frac{(R^2)^{\nu+1}}{4g\nu}[1+\alpha^2 - \frac{J^2}{R^4 \tilde
\phi^2}] \vec s_i =  \sum_{j(\ne i)} \frac{\vec 
s_{ij}}{(s_{ij}^2)^{\nu+1}}. \label{eqsi}
\end{equation}
Taking scalar product with $\vec s_i$ and dividing both sides by
$s_i^2 ~(\ne 0)$, we get, 
\begin{equation}
\frac{(R^2)^{\nu+1}}{4g\nu}[1+\alpha^2 - \frac{J^2}{R^4 \tilde
\phi^2}] =  \sum_{j(\ne i)} \frac{1 -(s_j/ s_i)
\cos(\theta_{ij})}{(s_i^2+s_j^2-2s_is_j\cos(\theta_{ij}))^{\nu+1}}.
\label{eqdot}  
\end{equation}
Note that the LHS is independent of the particle index $i$. Thus we have $N-1$
independent of equations of the type   
\begin{eqnarray}
&\sum_{j(\ne i)}& \frac{1 -(s_j/ s_i)
\cos(\theta_{ij})}{(s_i^2+s_j^2-2s_is_j\cos(\theta_{ij}))^{\nu+1}} 
\nonumber \\
&=&\sum_{j(\ne k)} \frac{1 -(s_j/ s_k)
\cos(\theta_{kj})}{(s_k^2+s_j^2-2s_ks_j\cos(\theta_{kj}))^{\nu+1}},
~~\forall ~~~k\ne i. \label{eqdt}
\end{eqnarray}
Further by taking the cross product with $\vec s_i$ and dividing by
$s_i$, we get, 
\begin{equation}
\sum_{j(\ne i)} \frac{s_j
\sin(\theta_{ij})}{(s_i^2+s_j^2-2s_is_j\cos(\theta_{ij}))^{\nu+1}} = 0,
\label{eqcrs}  
\end{equation}
which provides a further set of $N$ conditions on the internal
coordinates $\vec s_i$.  Notice that these conditions are manifestly
scale invariant. Together
eqs.(\ref{eqdt},\ref{eqcrs}) provide the $2N-1$ necessary
equations for determining the $s_i$ and the angles $\theta_i$. Notice that these determining 
equations are completely independent of $\alpha, J$ and $g$. We have
therefore the result that $( s_1, s_2,...,s_{N-1}, \theta_1,
\theta_2,...,\theta_{N})$ are independent of the magnetic field
($\alpha$), the 
total angular momentum $J$ and the interaction strength $g$. These
parameters, however, determine the overall scale $R$ through
eq.(\ref{eqdot}).  In fact since $s_N^2 =1$, the corresponding equation
may be taken to be the determining equation for $R$ in terms of the
parameters of the Hamiltonian,
\begin{eqnarray}
\frac{(R^2)^{\nu+1}}{4g\nu}[1+\alpha^2 - \frac{J^2}{R^4 \tilde
\phi^2}] ~~ = ~~~~~~~~~~~~~~~~~\nonumber \\
\sum_{j=1}^{N-1} \frac{1
-s_j\cos(\theta_{Nj})}{(1+s_j^2-2s_j\cos(\theta_{Nj}))^{\nu+1}} \equiv A(\nu,N)
\label{eqscale}  
\end{eqnarray}
Note that $A(\nu, N)$ introduced above is a function of 
$\nu$ and $N$ only. 
Thus we have a {\it very general }
result that the {\it geometry } 
or the shell structure of the equilibrium configuration is {\it independent} of
the parameters of the Hamiltonian which only restrict the overall size
of the system. The shell structure, however, depends on the nature of
repulsive interaction through the parameter $\nu$ but not its
strength. The above analysis  
is valid even if any one of the $s_i=0$, that is one particle being at
the origin of the coordinate system in which case we have two equations
less ( not more than one particle can be
at the origin). 

The energy of the equilibrium configuration can be easily computed
by noting that the auxiliary variable $\phi$ defined in
eq.(\ref{phidef}) is related to the two body potential energy by,
\begin{equation}
(1+\alpha^2-\frac{J^2}{\phi^2})\phi=  2g\nu  \sum_{i,j(\ne i)} \frac{1}
{(r_{ij}^2)^{\nu}}, \label{eqphi}
\end{equation}
where the RHS is proportional to the potential energy due to
interaction. Since the RHS and $\phi$ are positive definite we have
the condition $(1+\alpha^2)\phi^2 > J^2$. Interestingly taking further
gradients of the above equation again yield conditions on $\phi$ which
violate the above inequality. Thus the minima described by the
equations (\ref{eqpx},\ref{eqpy}, \ref{eqr}) must be {\it isolated}. 
We have for the energy at the extrema,
\begin{eqnarray}
E &=& \frac{1}{2}[ (1+\alpha^2)\phi + 2\alpha J + \frac{J^2}{\phi}] +
g\sum_{i,j(\ne i)} \frac{1}{(r_{ij}^2)^{\nu}} \nonumber \\
&=&
\frac{\nu+1}{2\nu}(1+\alpha^2)\phi + \alpha J +
\frac{\nu-1}{2\nu}\frac{J^2}{\phi}, \label{eqen}
\end{eqnarray}
where $\phi = R^2\tilde \phi$ and $\tilde \phi$ is independent of
$J$. This then is the energy of the equilibrium configuration in a
given $J$ sector. For the logarithmic case, the first line of
eq.(\ref{eqen}) has the second term corresponding to the logarithmic
interaction and the second line is not valid. 

In order to find the global minimum we
now minimise the energy with respect to $J$ and set $\partial
E/\partial J =0$, that is, 
\begin{eqnarray}
&&[(\nu+1)(1+\alpha^2)R -
(\nu-1)\frac{J^2}{\tilde\phi^2 R^3}]\frac{\partial R}{\partial
J} \nonumber \\
&+& \frac{\nu\alpha}{\tilde \phi} +
(\nu-1)\frac{J}{\tilde\phi^2 R^2} =0. 
\end{eqnarray}
Differentiating the $R$ w.r.t. $J$ in eq.(\ref{eqscale}) yields
$\partial R/ \partial J$ which when substituted in the above equation
gives the $J$ value and the corresponding global minimum energy,
\begin{equation}
J = -\alpha \tilde \phi R^2; ~~~E  = \frac{\nu+1}{2\nu}\tilde \phi R^2 ,
\end{equation}
where
\begin{equation}
R^2 = [ 4 g \nu A(\nu,N)]^{\frac{1}{\nu+1}} ,
\end{equation}
and $A(\nu, N)$ is defined before in eq.(\ref{eqscale}).
An important point to note here is that the minimum energy $E$ is 
{\it independent}
of the magnetic field and its dependence on $g$ is explicit. The
dependence on $N$ and $\nu$ is however involved. The angular momentum
$J$ at minimum of energy depends on the magnetic field and is zero in
the absence of the magnetic field as it should be. The 
expressions given above, though, are valid independent  of the
geometry of the clusters and are exact ( for approximate solutions
see eqs.(8,9) in ref.\cite{maksym} for the special case of Coulomb
interaction).   

The geometry of the clusters or shells are dependent on  $\tilde \phi$ and
$A(\nu,N)$ which are as yet unspecified. In general the equations for
the equilibrium configurations admit many solutions (which are
isolated as remarked earlier) for a given $N$ and $\nu$. 
In particular there are two special configurations which are always
solutions viz (i) all the $N$ particles are on a circle , $\bigcirc$ and
(ii) $N-1$ particles are on the circle with one particle at the center,
$\bigodot$ . For these two cases only the over all scale factor $R$ is to be
determined. The angles, $\theta_{ij}/2$, are simply multiples of $\pi/N$ and $\pi/(N -1)$
respectively.
These however need 
not be minimum energy configurations for a given $N$ and $\nu$. In fact
it has been 
numerically proved that for $N\le 5$ the circle configuration is
indeed the minimum energy 
configuration where as for $6\le N\le 8$ it is the circle-dot
which is the minimum
energy configuration in the case of Coulomb interaction ( $\nu=1/2$ ). 
Multiple shells start forming for $N\ge 9$. In what follows
we prove analytically that the first transition which occurs for $N$
from 5 to 6 is independent of $\nu$. 
The case of
circle and 
circle-dot is particularly simple since there is only one
scale involved. That is all $s_i^2 = 1, i=2,...,N$ and $s_1^2 =1$ for
the circle and $s_1^2 =0$ for the circle-dot. 

For the circle case, we have,
\begin{equation}
\tilde \phi = [\sum_{i=1}^{N-1} s_i^2 +1] = N
\end{equation}
and therefore the energy is given by, 
\begin{equation}
E_{\bigcirc}  =\frac{\nu+1}{2\nu}[ 4 g \nu A_{\bigcirc}
N^{\nu+1}]^{\frac{1}{\nu+1}} ,
\end{equation}
where 
\begin{equation}
A_{\bigcirc}(\nu,N) = \frac{1}{2^{2\nu+1}} \sum_{k=1}^{N-1}
\frac{1}{\sin^{2\nu}(\frac{k\pi}{N})}. 
\end{equation}

In the case of circle-dot, we have, 
\begin{equation}
\tilde \phi = N-1
\end{equation}
since there are now $N-1$ particles on the circle 
and therefore the energy is given by, 
\begin{equation}
E_{\bigodot}  =\frac{\nu+1}{2\nu}[ 4 g \nu A_{\bigodot}
(N-1)^{\nu+1}]^{\frac{1}{\nu+1}} ,
\end{equation}
where 
\begin{equation}
A_{\bigodot}(\nu,N) = A_{\bigcirc}(\nu,N-1)+1. 
\end{equation}
The extra 1 on the RHS is due to the contribution of the particle at the
center.
To ascertain which of these two configurations $\bigcirc$ and
$\bigodot$ has lower energy it is sufficient to look at the ratio,
\begin{equation}
(\frac{E_{\bigcirc}}{E_{\bigodot}})^{\nu+1} \equiv 
f(\nu,N) = (\frac{N}{N-1})^{\nu+1}
\frac{\lambda_N^{(\nu)}}{\lambda_{N-1}^{(\nu)}+2^{2\nu+1}} ,
\end{equation}
where,
\begin{equation} 
\lambda_N^{(\nu)} \equiv \sum_{k=1}^{N-1} \frac{1}{\sin^{2\nu}(\frac{k\pi}{N})}. 
\end{equation}
Note that in general the ratio $f$
depends only on $N$ and $\nu$. Obviously the circle is a lower energy
configuration iff $f < 1$. We claim that, for all $~ \nu ~ > ~ 0$, 
\begin{equation}
\begin{array}{lclr}
f(\nu,N) & < & 1  & ~~~ \mbox{for} ~~ N\le 5; \\
f(\nu,N) & > & 1  & ~~~ \mbox{for} ~~ N\ge 6. 
\end{array}
\end{equation}
Further the function $f(\nu,N)$ crosses unity exactly once for $N$
between 5 and 6 and nowhere else. This result can be easily seen for
$\nu=1$ since in this case $\lambda_N^{(1)} = (N^2-1)/3$. Therefore,
\begin{equation}
f(1,N)= \frac{N^2(N+1)}{(N-1)(N^2-2N+24)}
\end{equation}
which reproduces the claims made above, for $\nu=1$. The general proof
of these claims for {\it all} $\nu$ is some what involved and will be published
elsewhere. We sketch the arguments for the case $\nu
<<1 $, where one may use the expansion  $a^{\nu}\approx 1+\nu \log(a)$.
Using this $\lambda_N$ may be written as,
\begin{equation} 
\lambda_N^{(\nu)} \approx N-1 -2\nu X_N,
\end{equation}
where 
\begin{equation}
X_N = \log[\prod_{k=1}^{N-1}\sin(\frac{k\pi}{N})] =\log[\frac{N}{2^{N-1}}],
\end{equation}
where we have used the identity
$\prod_{k=1}^{N-1}\sin(\frac{k\pi}{N}) =\frac{N}{2^{N-1}}$. 
Substituting for $\lambda_N$ in $f$, we have
\begin{equation}
f(\nu,N) \approx
1+\frac{\nu}{N(N-1)}\mu_N, 
\end{equation}
where
\begin{equation}
\mu_N=[N(N-3)\log(N)-(N-1)(N-2)\log(N-1)]. 
\end{equation}
is independent of $\nu$. 
It is now easy to see that $\mu_N$ is negative for $N \le 5$ and positive
otherwise. Hence the claim. In the case of logarithmic interaction, the transition can
be seen more easily by taking the difference of energies for circle
and circle-dot since this
difference is independent of the arbitrary scale $\rho$ in the
interaction. It turns out that the difference is precisely given by
$g\mu_N$. Therefore the first geometric transition also occurs for the
log case exactly as in the power-law case. 

To summarize, we have proved in general that the organisation of many
body clusters in two dimensions into shells is a robust phenomenon
independent of the nature of the repulsive two-body interaction and also
independent of the Hamiltonian parameters but dependent only on the
number of particles in the cluster. In particular
we have analytically proved that the first geometric transition for
the ground state  from
circle to circle-dot configuration occurs after $N=5$.  
The robustness
of this transition seems to emerge purely from the number theoretic
properties of the ratios of the energies (difference in the log case)
in these two configurations. It is an open question if this is due to
some hidden symmetry properties. 
We have also
done numerical simulations for larger $N$ and for various values of
$\nu$. We find that the shell structure found by Bedanov and
Peeters\cite{bedanov} in the case of Coulomb interaction is valid for
all $\nu$ in general except that for larger $\nu$ the shell
description is valid only approximately. 
The equilibrium
configurations correspond to isolated minima. This fact should be
useful in calculating quantum fluctuations about the minima. The
details of these investigations will be published elsewhere. 

We thank R. Balasubramanian and D. Surya Ramana for help with number
theoretic identities. We also thank Madan Rao and Surajit Sengupta for
helpful discussions and for a critical reading of the manuscript.

\end{document}